\documentstyle[11pt]{article}

\input{epsfig.sty}

\def\boxit#1{
\vbox{\hrule height0.5st\hbox{\vrule width0.5st\kern10st\vbox{
\kern10st#1\kern10st}
\kern10st\vrule width0.5st}\hrule height0.5st}}

\def\bild#1\over#2{\mathrel{\mathop{\kern5st #1}\limits_{#2}}}

\newcommand{\be}{\begin{equation}}
\newcommand{\ee}{\end{equation}}
\newcommand{\bc}{\begin{center}}
\newcommand{\ec}{\end{center}}
\newcommand{\ba}{\begin{array}}
\newcommand{\ea}{\end{array}}

\newcommand{\vx}{\vec{x}}

\newcommand{\vy}{\vec{y}}
\newcommand{\vz}{\vec{z}}

\newcommand{\bphi}{\bar{\phi}}

\newcommand{\bx}{ {\bf x}}

\newcommand{\lap}{$\lambda \phi^4$ }

\begin{document}
\begin{titlepage}
\title{
 Space-time evolution of the \lap model:  classical 
and quantum aspects } 
\author{  F\'abio L. Braghin\thanks{ email:braghin@if.usp.br}  
and Fernando S. Navarra\thanks{ email:navarra@if.usp.br}  \\
{\normalsize Instituto de F\'\i sica da Universidade de S\~ao Paulo} \\
{\normalsize C.P. 66.318,  C.E.P. 05315-970, S\~ao Paulo,      
Brasil }
}
\maketitle
\begin{abstract}
A time dependent variational approach is used to derive 
the equations of motion for the \lap model. 
The simultaneous evolution 
of the quantum fluctuations  and 
of the classical part of the field 
is considered in a lattice of 1+1 
dimensions. Different initial conditions 
corresponding to non equilibrium
situations are considered and evolved in time.
Such high energy localized configurations expand in the 
lattice by ``bumps'' which may change with time. 
The quantum
fluctuations make the peaks be smoother and 
the expansion faster. 
\end{abstract}



 IFUSP- /2000.

\end{titlepage}

\section{ Introduction }

In relativistic heavy ion collisions, such as those performed now at 
RHIC and CERN, 
very dense and nearly baryon free systems are formed in the central rapidity 
region. These excited  systems decay and produce a large number of secondary
particles. In view of the large densities and  multiplicities, with or without 
quark gluon plasma formation, it is very likely that these intermediate systems 
will exhibit some collective behavior. It is therefore reasonable to treat their 
expansion with hydrodynamics \cite{dum,grassi,fredie} and it is
 interesting to investigate the connection
between the hydro-dynamical and the field theoretical approaches,
see for example \cite{EBOJAPI}.
In fact, the fluid approach has been taken so seriously 
that irregularities in the hydrodynamic flow pattern have been proposed as possible 
signals of a deconfined phase \cite{QM99}. 

Given the physical appeal of hydrodynamics, many of its aspects and underlying assumptions 
have been investigated during the last ten years. 
Some of the studied topics are thermalization,  initial conditions for the fluid expansion, 
the equation of state (and possible phase transitions), the freeze out mechanism and final 
state interactions. 
The direct correspondence of the hydro-dynamical description with the 
field theoretical one is thus highly appropriate.

In this context, at least two questions arise which  have not been addressed so far
 and are discussed in the present article. The first
is: i) which is the  role played by the quantum field nature  of this dense matter in the 
expansion?  In other words, which mistake are we doing when we use classical hydrodynamics 
and neglect quantum effects? The second question is: ii) if the expanding matter is merged 
in a condensate, which is the effect of this condensate on the expansion?

Whereas solving these equations is technically very complicated in full hydrodynamics, the 
interplay between  classical background and quantum fluctuations can be studied  
in simpler scenarios. In this work we discuss the space-time evolution of 
the \lap model in $1+1$ dimensions at zero temperature. Solving nonperturbatively 
and self-consistently 
the equations of motion of the theory,  we can follow the expansion of a highly energetic 
system composed by a self interacting scalar field. In particular, we can separately investigate the 
classical and quantum components of the system and estimate the  role played   by  
the condensate during the expansion.

In spite of the enormous differences
between this simple model 
and the realistic descriptions of the 
 ``fireballs'' formed at 
RHIC, we believe that our study  
can be of interest and give us some 
relevant insights into the real physical situation.

Due to the extreme complexity of realistic theories, such  as QCD, 
one usually is obliged to consider effective models which respect the major 
properties of the fundamental theory. The \lap model shares with QCD the 
properties of asymptotic freedom and spontaneous symmetry breaking (SSB) 
in the frame
of the Gaussian approximation. Due to the SSB, the field is decomposed 
into a classical and a quantum part.
At the 
same time it may also represent the mesonic sector of the linear sigma 
model, which is often used to describe the dynamics of a gas 
of pions.  In the context of cosmology the 
scalar field of the model  may be considered as the relevant degree of freedom 
for inflationary models \cite{COSMO}.

Some progress has been achieved in the last 
decade in understanding the dynamics of the \lap model. In particular, lattice 
calculations of the time evolution of  out of equilibrium configurations, both 
at  zero and finite temperatures, in 1+1 and 3+1 dimensions
in continuous and discretized space-times have been performed
\cite{COOPER,BACKE,FLB98a,BOYA,ENCONTROS,SSV}.

In this work the \lap will be investigated in  
the Gaussian approach with the formalism discussed in Refs. 
\cite{FLB98a}. It is suitable for non homogeneous configurations of the
fluctuations (the quantum part of the field) as well as of the condensate 
(as the classical part of the field $\bphi(t)$ will be called).  
They can be thought of
as two interacting ``liquids'' with continuous energy transfer and this view is  
under investigation.
The text is organized as follows. In section II the Gaussian  formalism for
out of equilibrium systems is briefly outlined and 
the equations of movement are derived. 
In section III we discuss the numerical method employed here
and developed in \cite{FLB98a}.
The numerical results are presented in section IV for several different 
out of equilibrium situations and values of the coupling constant.
The results and some perspectives are summarized in the final section.

\section{ Time dependent Gaussian approximation to $\lambda \phi^4$  model }

The Hamiltonian density for a scalar field   $\phi$ with bare mass   $m_0^2$
and coupling constant  $\lambda$ is: 
\begin{eqnarray} \label{2}
 H = \frac{1}{2} \left( \pi^{2}(\vx) +  (\nabla
\phi (\vx))^2 +  {m_0^2} \phi^{2}(\vx) + \frac{\lambda}{12}\phi^{4}(\vx) \right),
\end{eqnarray}
where the action of operators  $\phi$ and  $\pi$ in functional Schroedinger 
representation over a wave functional 
$\Psi \left[ \phi (\vx) \right] = < \phi (\bx) | \Psi \left[\phi \right] > $
is given by:
\be \label{2a} \ba{ll}
\displaystyle{ \hat{\phi} |\Psi \left[ \phi(\vx) \right] >
= \phi(\vx) |\Psi \left[\phi(\vx) \right] >,}\\
\displaystyle{ \hat{\pi} |\Psi \left[\phi (\vx)\right] >  =
- i \delta / \delta \phi (\vx) |\Psi \left[\phi(\vx) \right] >.  }
\ea
\ee
For our variational calculations  analyzed below,  in the Schroedinger  picture the wave 
functional evolves like the Schroedinger 
equation  
\be \label{3}
\displaystyle{ i \frac{ \partial}{\partial t} \Psi \left[\phi (\vx)\right] =
H \Psi \left[\phi (\vx)\right]. }
\ee
This is, thus, a non covariant formalism suitable for time dependent
problems.

In the Gaussian approximation at zero temperature  
$\Psi$ is parametrized by:
\be \label{4} \ba{ll}
\displaystyle{
\Psi\left[\phi(\vx )\right] = N  exp \left\{ -\frac{1}{4} \int
d \vx d \vy \delta\phi(\vx )\left(G^{-1}(\vx ,\vy) + i\Sigma(\vx,\vy)
\right) \delta \phi(\vy)
+ i\int d\vx \bar{\pi} (\vx)\delta \phi (\vx) \right\} ,  }
\ea
\ee
Where 
$ \delta\phi(\vx, t) = \phi(\vx)-\bar\phi(\vx,t) $; 
 the normalization is  $N$, the variational parameters are   
the condensate $ \bar \phi (\vx , t) = < \Psi | \phi | \Psi > $ and its 
conjugated variable $ \bar{\pi}(\vx, t) = < \Psi | \pi | \Psi > $;
quantum fluctuations are represented by the width  of the Gaussian  
$ G(\vx,\vy,t) = <\Psi |\phi(\vx) \phi(\vy)  | \Psi> $ and its conjugate
variable $ \Sigma(\vx,\vy,t) $.

In variational time dependent calculations  we have to choose an action 
to be  minimized in order  to obtain the equations of motion.
 We take the 
well known Dirac action from  \cite{JACKER}:
\be \label{5}
\displaystyle{
I = \int dt <\Psi \mid \left( i \frac{\partial}{\partial t} - \hat H
\right) \mid \Psi> .  }
\ee
In order to calculate it, we take the mean value of an operator   $\hat{O}$ 
given by:
\be \ba{ll}
\displaystyle{ < \Psi | \hat{O} | \Psi > = \int {\cal D} \left[ \phi \right]
\Psi^* \hat{O} \Psi   }
\ea
\ee
By means of this average procedure the energy density $\rho = <\Psi|H|\Psi>$ 
can be calculated. Its temporal evolution will be investigated in section 4.

Variations with respect to the variational parameters and their 
conjugated yield the following  equations of motion
(any repeated  spatial index means integration 
over that variable):
\be \label{7} \ba{ll}
\displaystyle{ \frac{\delta I}{\delta \Sigma(\vx,\vy) } \rightarrow
\partial_t G(\vx,\vy) = 2  \left( G(\vx,\vz)\Sigma(\vz,\vy) +
 \Sigma(\vx,\vz)G(\vz,\vy) \right) , }  \\
\displaystyle{ \frac{\delta I}{\delta G(\vx,\vy)} \rightarrow
\partial_t \Sigma(\vx,\vy) =  \left(
2 \Sigma(\vx,\vz)\Sigma(\vz,\vy) -
\frac{1}{8}G^{-1}(\vx,\vz)G^{-1}(\vz,\vy) \right)  +
 \left( \frac{\Gamma(\vx,\vy)}{2} +
\frac{ \lambda}{2} \bar \phi(\vx)^2 \right),  } \\
\displaystyle{
\frac{\delta I}{\delta \bar{\pi}(\vx)} \rightarrow
\partial_t \bar \phi(\vx) = - \bar{\pi} (\vx),  }  \\
\displaystyle{
\frac{\delta I}{\delta \bar{\phi}(\vx) } \rightarrow
\partial_t \bar{\pi}(\vx) = \Gamma(\vx,\vy)\bar \phi(\vy) +
\frac{ b}{6}\bar \phi^2(\vx), }
\ea
\ee
Where
$\Gamma (\vx,\vy) = -\Delta  + \left( m_0^2 + \frac{ \lambda }{2}G(\vx ,\vx )
\right)\delta (\vx-\vy)$.
In this  approximation  the interaction term $ \lambda \phi^4$  becomes 
quadratic, i.e., it contributes to a self consistent mass.
 These equations were generalized for the out of thermo-dynamical equilibrium 
using different methods in \cite{EBOJAPI,COOPER}.  

Using the Gaussian ansatz (\ref{3})
in the symmetric phase ($\bar{\phi}=0$) we need only two initial 
conditions for the temporal evolution of these equations,
$G(t=0)$ and  $\dot{G}(t=0)$, which is proportional to its imaginary 
part $\Sigma$. 
An inspection of the equations of motion shows that for $\bphi(t=0) = \dot{\bphi} (t=0) =0$
the variables $\bphi$ and $\bar{\pi}$ will not change, i.e.,  the classical part of the field will 
remain constant (zero) during all times in the frame of this approach.

Several
numerical works have been done for the above equations of movement.  The choice of initial conditions
is entirely subordinate to the Gaussian approximation. 
If it were not
Gaussian, we would have to consider three conditions instead of
two \cite{COOMOT}.  The analysis of  
these equations by Boyanovsky {\it et al} and by \cite{FLB98a} shows that initial conditions 
(for homogeneous $G$ and $\bphi$ ) are  crucial for the time interval 
in which the system evolves towards the minimum and 
for the speed of the field evolution.

\section{ Numerical method}

The numerical method for the temporal evolution of the initial conditions 
corresponding to the system described above used in this paper was 
developed in \cite{FLB98a,BLARIP}. 
This section is a brief review.
One defines a generalized density matrix in a lattice as:
\begin{eqnarray}
     R_{i,j} =   \left(
        \begin{array}{ll}
          \rho_{i,j}  &  \kappa_{i,j}   \\
       -\kappa^{\ast}_{i,j}  & -\rho^{\ast}_{i,j}
         \end{array}
        \right),
\end{eqnarray}
where the density matrices (mean values) are given by:
$\rho_{i,j} = \frac{1}{2}<a_ia^{\dag}_j
+ a^{\dag}_i a_j > $ which is hermitian and 
$ \kappa_{i,j}= -<a_i a_j > $ is symmetric, using the creation and 
annihilation operators. These operators can be written in a lattice with mesh size
 $\Delta x$ in d spatial dimensions as:
\begin{eqnarray}
a(j) & = & \frac{1}{\sqrt{2}}\left\{ \phi(j)(\Delta x)^{\frac{d-1}{2}} +
   i \pi(j)(\Delta x)^{\frac{1+d}{2}} \right\} \\
a^{\dag}(j) & = & \frac{1}{\sqrt{2}}\left\{ \phi(j)
(\Delta x)^{\frac{d-1}{2}} -
   i \pi(j)(\Delta x)^{\frac{1+d}{2}} \right\}
\end{eqnarray}
Some of the calculated mean values in terms of the matrix elements of the 
above matrix  are given by:
\be \label{63} \ba{ll}
\displaystyle{ G(i,j) = <\phi(i) \phi(j) > = \frac{1}{(\Delta x)^{d-1} }
\Re e ( \rho(i,j) - \kappa(i,j) )   }\\
\displaystyle{ F(i,j) = <\Pi(i) \Pi(j) > = \frac{1}{(\Delta x)^{d+1} }
\Re e ( \rho(i,j) + \kappa(i,j) ) }
\ea
\ee
where $ F(i,j) = G(i,j)^{-1}/4 + 4 \Sigma(i,k)  G(k,l) \Sigma(l,i) $.

The temporal evolution is governed by the Hartree (Fock) Bogoliubov energy, 
which can be parametrized in the following form:
\begin{eqnarray}
  \frac{1}{2} H_{i,j} = \frac{\delta E}{\delta R_{j,i} }=
          \left(
                     \begin{array}{cc}
                    W_{i,j}  & D_{i,j} \\
                   -D_{i,j} & -W_{i,j}
                      \end{array}
                  \right),
\end{eqnarray}
where the above matrices are given
in terms of the parameters of the model \cite{FLB98a}:
\begin{eqnarray}
W_{i,j}= \frac{1}{2}( -\Delta_{i,j} + (m_0^2 + \lambda G_{i,i}/2 +
(\Delta x)^{-2} )\delta_{i,j}), \\
D_{i,j}= \frac{1}{2}( -\Delta_{i,j} + (m_0^2 + \lambda G_{i,i}/2 - 
(\Delta x)^{-2} )\delta_{i,j}). 
\end{eqnarray}
With these matrices, one can check that the Liouville-von 
Neumann condition is satisfied:
\be
i \dot R_{i j} = \left[ H_{i k} , R_{k j} \right].
\ee
These equations are equivalent to those obtained from the 
Gaussian approximation of the previous section \cite{FLB98a}.
The time-evolution of the generalized density matrix can be then performed for 
given initial conditions. 

\section{ Numerical results}

In this section we will report results of numerical calculations for the
equations of motion studied above in a lattice of 100 points with spacing $\Delta x = 0.1fm$.
By varying  $\Delta x \to 0$ it was possible to assure the reliability of the results (dynamics does
not change in this limit).  The lattice spacing 
 is always to be much smaller than the correlation length 
($\xi = 1/m_{phys}$) which is quite smaller than the lattice size (10 fm).
The first initial condition for the
classical part of the field is given by:
\be \label{V2}
\bphi (t=0) = \bphi_0 \,\, tanh^2 \left( \frac{x - x_0}{0.5} \right), 
\;\;\;\;\;
\;\;\;\;\;  \dot{\bar{\phi}} = {\bar{\pi}} = 0,
\ee
where $x_0$ is the center of the lattice and the coordinate $x$ is discretized. 
This configuration is not stable since in a small region of the lattice
the field has values different from the vacuum. 

The field configuration corresponding to (\ref{V2}) is shown in 
Figure 1 with a solid line for the model with a 
strong coupling constant $\lambda = 600 fm^{-2}$ and physical
mass $\mu = 100 MeV$. In terms of 
energy it corresponds to a bubble of high energy density  in the vacuum which
can be interpreted as an {\it in medium} effect. 
This initial condition is plugged into the system of equations (\ref{7}) and numerically 
evolved with the method described in the last section. Knowing the field configuration  
at all times we can compute the energy density distribution at all  time steps for 
different scenarios. Periodic boundary conditions are considered in this work.

In Figure 2a  we show the energy density distribution 
of the classical part of the field alone (which we call $\rho$)
along the $x$ direction   for different 
times $t = 0$ (thick solid),  $=0.1$ (solid), $=1$ (thin solid), $2$ (thick dotted) 
and $3.5$ (dotted) fm. As it can be seen, the energy density excess with relation to the 
vacuum value is distributed among the 
lattice by means of ``waves''. 
The expansion of the initially high energy density proceeds
with two peaks for each side of the lattice. This feature may be expected
for the massless Klein-Gordon equation which has a sinoidal solution
due to the gradient terms. The energy density is proportional to the square 
of the field and consequently we obtain this ``two bumps'' structure.

In Figure 2b the same initial conditions for the condensate
 are evolved in time with the 
contribution from the quantum fluctuations, i.e., considering all the four equations (\ref{7}). 
For the quantum part along this work, it was assumed that, at the initial time,
they are at the vacuum value, i.e.:
\be \label{V3a}
G(r,t=0) = G_0  \;\;\;\;\;\;\;\;\;\;  \Sigma \propto \dot{G}(r,t=0) = 0,
\ee
where $G_0$ is the value of the quantum fluctuations in the vacuum of the
asymmetric phase ($\bphi \neq 0$) obtained from a gap equation. 
In this case the gap equation is given by:
\be
\mu^2 = m_0^2 + \frac{\lambda}{2} G_0(\mu^2) + \frac{\lambda}{2} \bphi_0^2
\ee
The value of $G_0$ can be  fixed by the physical mass $\mu$. 
There is, now, energy transfer 
from the classical part to the quantum fluctuations and back.  
Comparing  figures 2a and 2b we can clearly see that the inclusion of 
quantum fluctuations accelerates the expansion. This is especially 
visible at $t = 3.5$ fm and even more at higher times. 
The effect is small but unambiguous. We observe an inversion in the heights of the two bumps 
that characterize the energy distribution, the outer bump becoming more 
pronounced in the quantum case.  
This is an indication that, quantum fluctuations 
increase the pressure in spite of the isoentropic character of the approximation.
Furthermore, the energy density configuration is clearly smoother than in the
purely classical case. 
It is worth to remember once more 
that the present approach is not
yet completely ``self-consistent'' in the sense that for  given  initial
``non-equilibrium'' conditions the model should reflect this fact including,
for instance, the coupling constants and quantum fluctuations.
This is under study \cite{FB2001}.

In order to investigate the sensitivity of the  conclusion found in the last 
paragraph to the shape of the initial conditions, we repeat the same calculations 
starting from the following field profile:
\be \label{V3} 
 \bphi (t=0) = \bphi_0  \left\{ 1  +  0.3 \,\, sech^2 
\left( \frac{x - x_0}{0.5} \right) \right\}. 
\ee
The corresponding field configuration  is shown in Figure 1 with a dashed line.
In this case we use a different coupling constant $\lambda = 60 fm^{-2}$. This 
choice  changes  the global normalization of the field and energy, but it does
not change the dynamics considerably. As a matter of fact, in the classical level,
there is a scale invariance in the dynamics with relation to changes in such variable.
Taking into account quantum fluctuations this scale invariance is broken (slightly
for the range of parameters considered in this work).
Evolving only the classical part of the field in the same way as before we obtain 
Figure 3a, which is very similar to Fig. 2a. Switching on the quantum fluctuations
leads to results of Fig. 3b. 
Comparing Figs. 3a and 3b we observe the same effect already seen 
in Fig. 2, i.e., the inclusion of quantum fluctuations accelerates the expansion.

In the asymmetric phase the effective potential has a Mexican hat like form and in all cases 
presented in this work the total energy is smaller than the ``barrier'' (located at the origin
of the $\bphi$ space)
 which separates the two  minima of the potential.
In the cases shown in Figures 2b and 3b there is a tunneling of the condensate from one side 
of the potential to the other. For some kinds of initial condition this was 
already shown  in \cite{ENCONTROS}.
This effect disappears if  the quantum fluctuations are switched off but remain when 
the lattice spacing is reduced as discussed below.

The presence of quantum fluctuations make the energy density to be amplified in the earlier times
and to expand faster. This amplification is stronger in the case shown in figure 3b.
As time goes the energy density ``waves'' tend to decrease.
Moreover, we notice small  regions with energy density 
lower than the true vacuum of the corresponding phase in the beginning of the expansion, just  
after the energy density amplification happens.

The total energy of the system is conserved during the evolution.
We want to stress that the inclusion of quantum fluctuations changes 
the overall energy normalization of the system. 
Therefore, the  initial energy profile is similar
for all figures but the normalization is different as can be seen in the pictures.

The  equations of movement (\ref{7})
exhibit Ultra Violet (UV) divergences in a continuum space-time which
are exactly the same as those present in the GAP equation for the vacuum of the model. They 
require mass renormalization (and also coupling constant renormalization depending on the space-time 
dimension) which is exactly the same for the static and time dependent cases. 
In \cite{FLB98a,ENCONTROS}  analytical solutions for these equations of
movement were found for a particular kind of initial conditions. They exhibit the UV divergences only 
at the initial time (t=0). As this divergence is eliminated by the renormalization 
(of the mass and/or coupling) the temporal evolution remains unchanged.
This is also true in the lattice: the deviation of the system with relation to the vacuum state
determines the temporal evolution.
For a variation of the lattice spacing, i.e., as $\Delta x \to 0$ (for low values as 
0.02 fm were checked) the dynamics does not change considerably (with the corresponding 
mass rescaling).
In fact, the continuum limit must be the same for the static and time-dependent cases. 
The deviations from this (renormalized) state are always finite and
are not  altered by vacuum redefinition (renormalization). 
Mass 
renormalization  changes the normalization of the energy density being, therefore, not important
for the dynamical evolution.

In fact, the difference between the classical system and the classical plus 
quantum  system  is, in part,  related to a scale invariance. The 
classical level exhibits scale invariance but this is not present at the quantum level.
 These effects will be explicitly 
addressed in detail  elsewhere  \cite{FB2001}.

Our results point out to the following simple picture: without quantum
fluctuations a given localized spatial configuration of the field $\bphi$ 
expands,  simply converting ``potential'' into ``kinetic'' energy. The inclusion 
of these fluctuations accelerates the expansion. This effect 
may be caused by the interaction between the particles and the 
condensate, by the tunneling between the two vacua or by these  both aspects.   
The questions addressed in this work are related to those discussed in
Ref. \cite{dum}.
In that work, the expansion was driven by the effective potential, (which 
contains a \lap part) plus the contribution of a gas of pions. The former 
and latter contributions correspond respectively to our classical and quantum 
contributions. The total pressure is the sum of both the effective potential and 
the pion gas component and therefore also in that model the inclusion of 
quantum fluctuations increases the total pressure and the expansion rate. 
These conclusions also are in agreement with studies for the effects
of quantum fluctuations in the Inflationary scenario where there is
``acceleration'' of the dynamics, see for example \cite{BOYA,GUPI}.

\section{Summary}

We have analyzed the temporal evolution of non-homogeneous configurations
of the \lap model considering two different approaches: the classical
equations of motion and the equations of 
motion in the framework of the Gaussian approach for the quantum 
fluctuations in a one-dimensional lattice. 
We have been able to study the influence of the quantum 
fluctuations on the classical field dynamics and vice versa. Quantum
fluctuations make the energy distribution smoother.
Besides that, we conclude that the quantum fluctuations 
accelerate the expansion of a non homogeneous configuration
of the classical part of the field after having amplified the energy density. 
We noted a  tunneling of the condensate from one side of the ``Mexican-hat'' potential to the other
when quantum fluctuations are considered depending on the initial conditions.

We would like to emphasize that in the limit of vanishing
lattice spacing the dynamics is not qualitatively modified: the 
renormalization acts as to change
the field and therefore energy density normalization but not the dynamics.
We have been concerned mainly with short interval evolution, but the 
evolution of 
 thermal degrees of freedom  will be analyzed elsewhere.
Finally with the inclusion of temperature, we hope to establish a 
closer connection between our system and a hydro-dynamically expanding 
fireball.

{\bf Acknowledgements}\\
This work was supported in part by FAPESP- Brazil. F.L.B. wishes to thank
D. Vautherin with whom part of this work was initiated.
Most part of the numerical calculations was performed in the machines of
the Laboratory for Computation of the University of S\~ao Paulo - LCCA-USP.

\newpage

\noindent
{\bf Figure Captions}\\
\begin{itemize}

\item[{\bf Fig. 1}] Initial field configuration. Solid line corresponds to 
(\ref{V2}) and the dashed line line to (\ref{V3}).

\item[{\bf Fig. 2}] a) Evolution of the energy distribution of the condensate 
at different times; b) the same as a) with the inclusion of quantum fluctuations.
The initial field configuration is given by (\ref{V2}).

\item[{\bf Fig. 3}] The same as Fig.2 for the initial configuration (\ref{V3}).

\end{itemize}

\newpage

\begin{figure}[htb]

\epsfig{figure=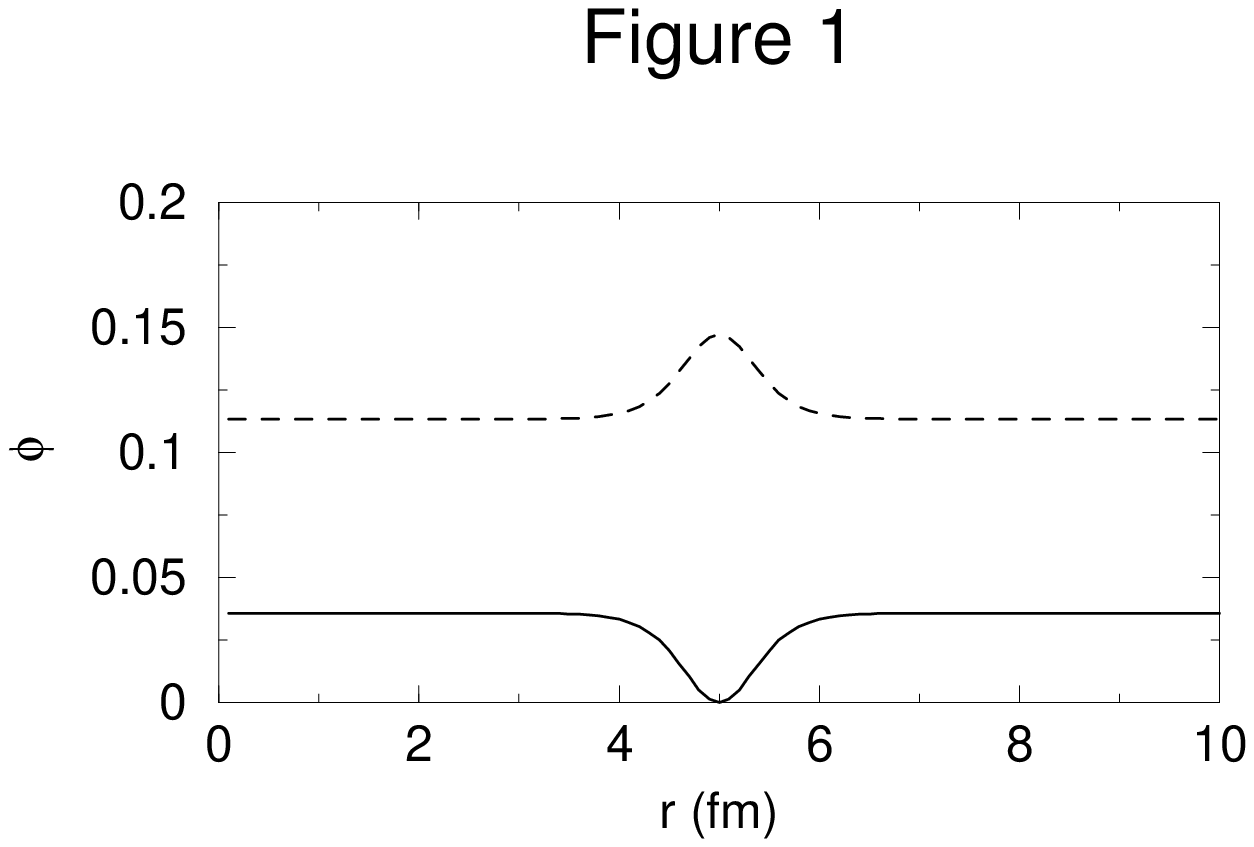,width=16cm} 

\end{figure}

\newpage

\begin{figure}[htb]
\epsfig{figure=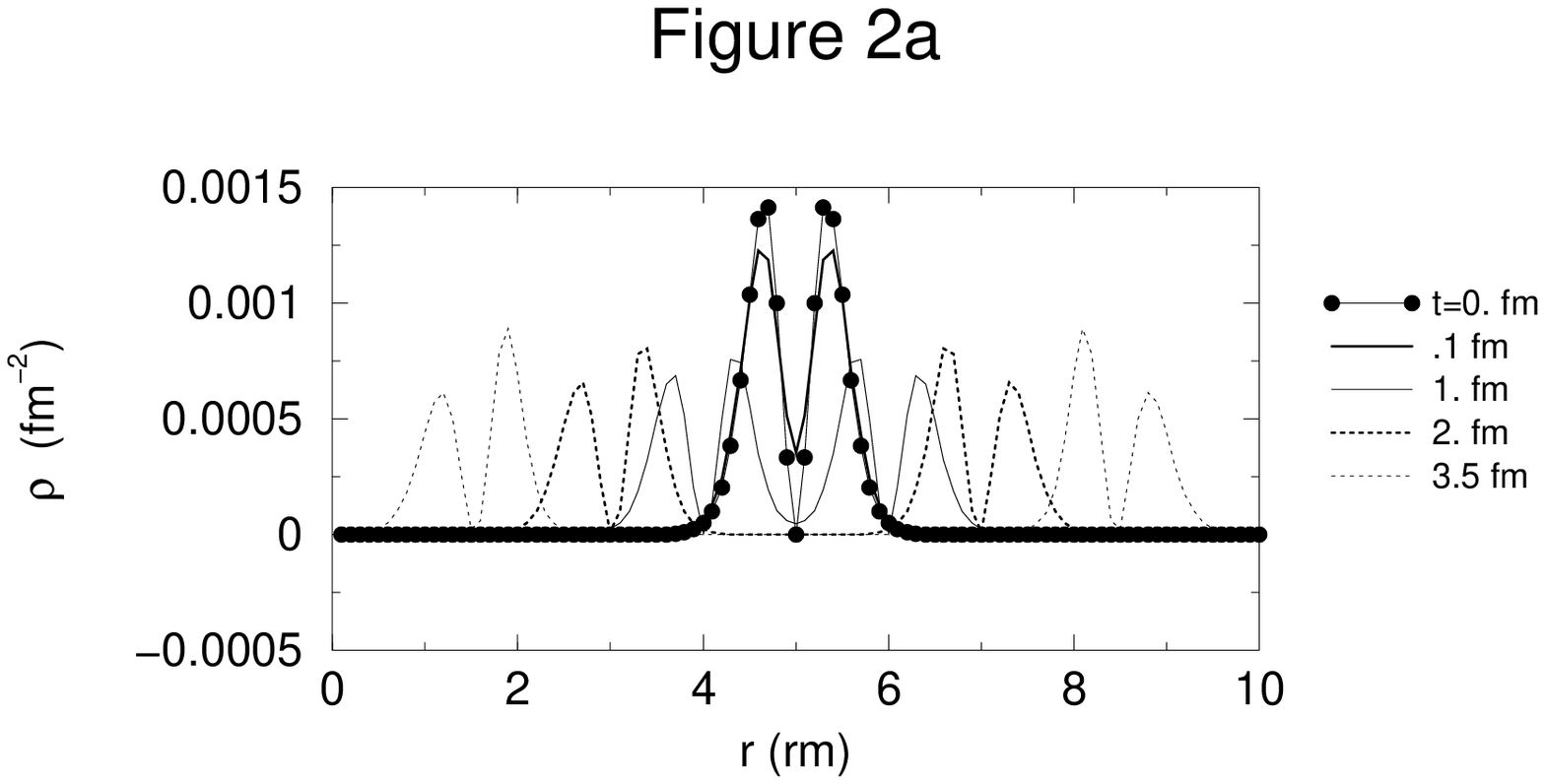,width=14cm} 

\epsfig{figure=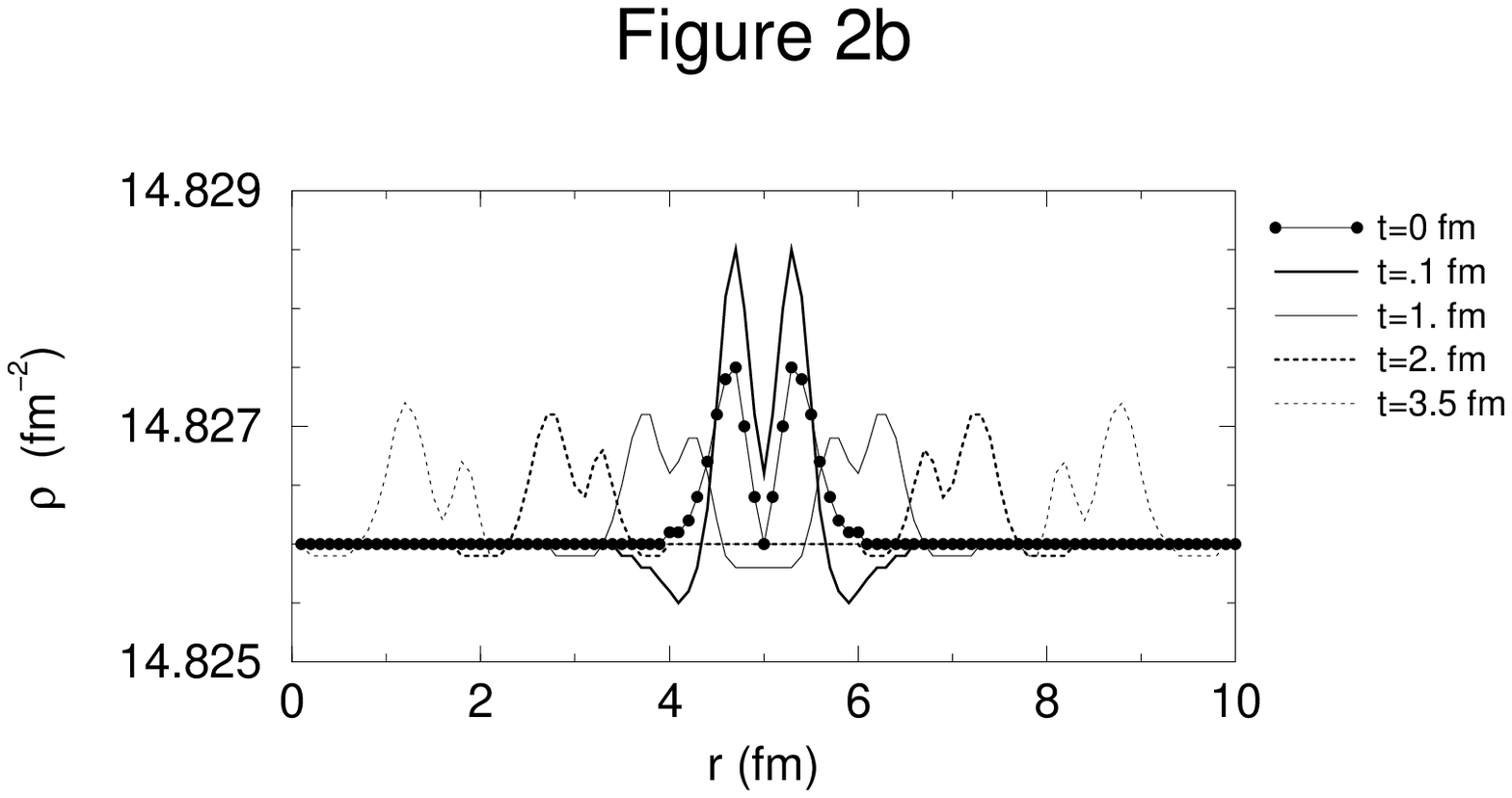,width=14cm}
\end{figure}

\newpage

\begin{figure}[htb]
\epsfig{figure=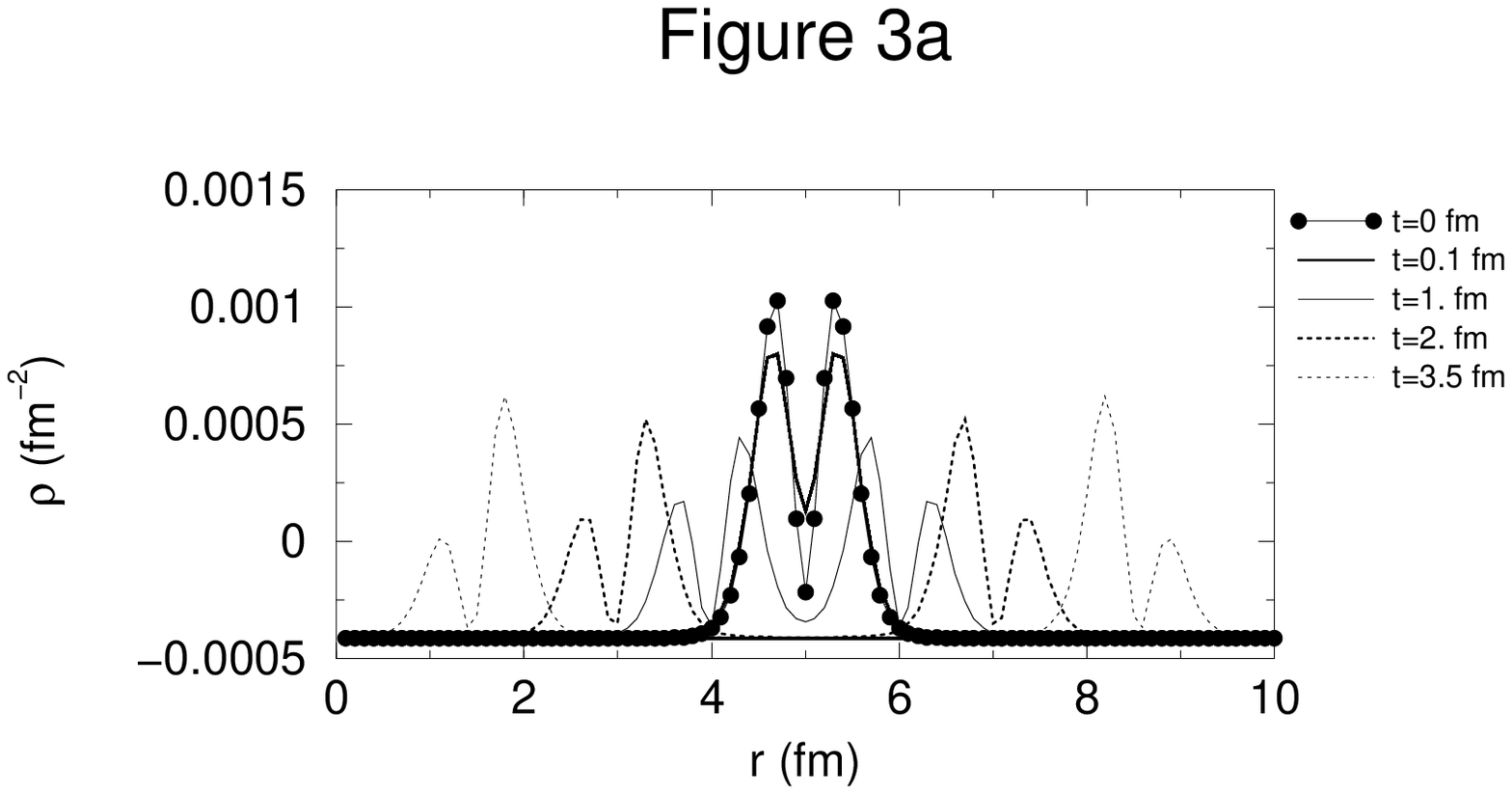,width=14cm} 

\epsfig{figure=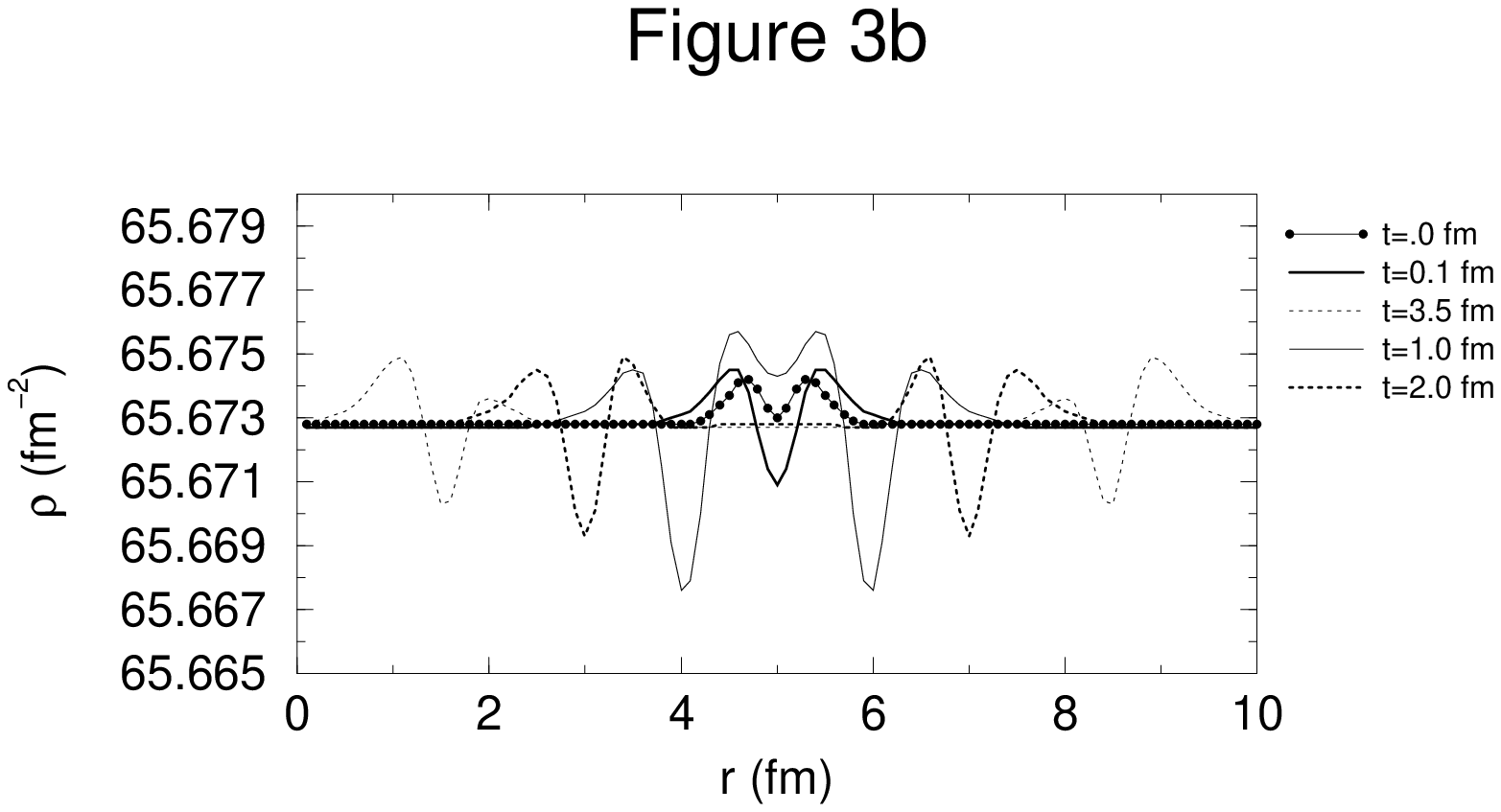,width=14cm}
\end{figure}

\end{document}